\documentstyle[12pt,epsf]{article}
\newcounter{sectionc}\newcounter{subsectionc}\newcounter{subsubsectionc}
\renewcommand{\section}[1] {\vspace{0.6cm}\addtocounter{sectionc}{1}
\setcounter{subsectionc}{0}\setcounter{subsubsectionc}{0}\noindent
	{\bf\thesectionc. #1}\par\vspace{0.4cm}}
\renewcommand{\subsection}[1] {\vspace{0.6cm}\addtocounter{subsectionc}{1}
	\setcounter{subsubsectionc}{0}\noindent
	{\it\thesectionc.\thesubsectionc. #1}\par\vspace{0.4cm}}
\renewcommand{\subsubsection}[1]
{\vspace{0.6cm}\addtocounter{subsubsectionc}{1}
	\noindent {\rm\thesectionc.\thesubsectionc.\thesubsubsectionc.
	#1}\par\vspace{0.4cm}}
\newcommand{\nonumsection}[1] {\vspace{0.6cm}\noindent{\bf #1}
	\par\vspace{0.4cm}}

\renewenvironment{thebibliography}[1]
	{\begin{list}{\arabic{enumi}.}
	{\usecounter{enumi}\setlength{\parsep}{0pt}
\setlength{\leftmargin 1.25cm}{\rightmargin 0pt}
	 \setlength{\itemsep}{0pt} \settowidth
	{\labelwidth}{#1.}\sloppy}}{\end{list}}

 \catcode`@=11
\def\@cite#1#2{\unskip\nobreak\relax
    \def\@tempa{$\m@th^{\hbox{\the\scriptfont0 #1}}$}%
    \futurelet\@tempc\@citexx}
\def\@citexx{\ifx.\@tempc\let\@tempd=\@citepunct\else
    \ifx,\@tempc\let\@tempd=\@citepunct\else
    \let\@tempd=\@tempa\fi\fi\@tempd}
\def\@citepunct{\@tempc\edef\@sf{\spacefactor=\the\spacefactor\relax}\@tempa
    \@sf\@gobble}

\def\citenum#1{{\def\@cite##1##2{##1}\cite{#1}}}
\def\citea#1{\@cite{#1}{}}

\newcount\@tempcntc
\def\@citex[#1]#2{\if@filesw\immediate\write\@auxout{\string\citation{#2}}\fi
  \@tempcnta\z@\@tempcntb\m@ne\def\@citea{}\@cite{\@for\@citeb:=#2\do
    {\@ifundefined
       {b@\@citeb}{\@citeo\@tempcntb\m@ne\@citea\def\@citea{,}{\bf ?}\@warning
       {Citation `\@citeb' on page \thepage \space undefined}}%
    {\setbox\z@\hbox{\global\@tempcntc0\csname b@\@citeb\endcsname\relax}%
     \ifnum\@tempcntc=\z@ \@citeo\@tempcntb\m@ne
       \@citea\def\@citea{,}\hbox{\csname b@\@citeb\endcsname}%
     \else
      \advance\@tempcntb\@ne
      \ifnum\@tempcntb=\@tempcntc
      \else\advance\@tempcntb\m@ne\@citeo
      \@tempcnta\@tempcntc\@tempcntb\@tempcntc\fi\fi}}\@citeo}{#1}}
\def\@citeo{\ifnum\@tempcnta>\@tempcntb\else\@citea\def\@citea{,}%
  \ifnum\@tempcnta=\@tempcntb\the\@tempcnta\else
   {\advance\@tempcnta\@ne\ifnum\@tempcnta=\@tempcntb \else \def\@citea{--}\fi
    \advance\@tempcnta\m@ne\the\@tempcnta\@citea\the\@tempcntb}\fi\fi}

\newcommand{\tcaption}[1]{
        \refstepcounter{table}
        \setbox\@tempboxa = \hbox{\tenrm Table~\thetable. #1}
        \ifdim \wd\@tempboxa > 6in
           {\begin{center}
        \parbox{6in}{\tenrm\baselineskip=12pt Table~\thetable. #1}
            \end{center}}
        \else
             {\begin{center}
             {\tenrm Table~\thetable. #1}
              \end{center}}
        \fi}
\catcode`@=12

\def\lsim{\mathrel{\raise.2ex\hbox{$<$}\hskip-.8em\lower.9ex\hbox{$\sim$}}}
\def\gsim{\mathrel{\raise.2ex\hbox{$>$}\hskip-.8em\lower.9ex\hbox{$\sim$}}}

\begin{document}

\parindent=1.5pc

\font\fortssbx=cmssbx10 scaled \magstep2
\hbox to \hsize{
\includegraphics{uwlogo.ps}
\hskip.5in \raise.1in\hbox{\fortssbx University of Wisconsin - Madison}
\hfill$\vcenter{\hbox{\bf MADPH-95-887}
                \hbox{June 1995}}$ }

\vspace{.25in}

\begin{center}
\bf
The Direct and Indirect Detection of\\ Weakly Interacting Dark Matter
Particles\\[3mm]
\small
F. Halzen\\[1mm]
\footnotesize\it
University of Wisconsin, Department of Physics, Madison WI 53706\\
and\\
University of Hawaii, Department of Physics, Honolulu HI 96822
\end{center}

\vspace{.3cm}

\begin{center}
\footnotesize
ABSTRACT
\medskip

\parbox{5in}{\baselineskip13pt\looseness=-1
An ever-increasing body of evidence suggests that weakly interacting massive
particles (WIMPs) constitute the bulk of the matter in the Universe.
Experimental data, dimensional analysis and Standard Model particle physics are
sufficient to evaluate and compare the performance of detectors searching for
such particles either directly (e.g.\ by their scattering in germanium
detectors), or indirectly (e.g.\ by observing their annihilation into neutrinos
in underground detectors). We conclude that the direct method is superior if
the WIMP interacts coherently and its mass is lower or comparable to the weak
boson mass. In all other cases, i.e.\ for relatively heavy WIMPs and for WIMPs
interacting incoherently, the indirect method will be competitive or superior,
but it is, of course, held hostage to the successful deployment of high energy
neutrino telescopes with effective area in the $\sim10^4$--$10^5$~m$^2$ range
and with appropriately low threshold. The rule of thumb is that a kilogram of
germanium is roughly equivalent to a $10^4$~m$^2$ neutrino telescope, although
the signal-to-noise is, at least theoretically, superior for the neutrino
detector. The energy resolution of the neutrino telescope may be exploited to
measure the WIMP mass and suppress the background. A kilometer-size detector
probes WIMP masses up to the TeV-range, beyond which they are excluded by
cosmological considerations.}

\end{center}

\thispagestyle{empty}

\section{Introduction and Results}

It has become widely accepted that most of our Universe is made of cold dark
matter particles. Big bang cosmology implies that these particles have
interactions of order the weak scale, i.e.\ they are WIMPs\cite{Seckel}. We
briefly review the argument which is sketched in Fig.~1. In the early Universe
WIMPs are in equilibrium with photons. When the Universe cools to temperatures
well below the mass $m_\chi$ of the WIMP their density is Boltzmann-suppressed
as $\exp(-m_\chi/T)$ and would, today, be exponentially small if it were not
for the expansion of the Universe. At some point, as a result of this
expansion, WIMPs drop out of equilibrium with other particles and a relic
abundance persists. The mechanism is analogous to nucleosynthesis where the
density of helium and other elements is determined by competition between the
rate of nuclear reactions and the expansion of the Universe.

At high temperatures WIMPS are abundant and they rapidly convert into lighter
particles. Also, as long as they are in equilibrium, lighter particles interact
and create WIMPs. The situation changes rapidly after the temperature drops
below the threshold for creating WIMPs, $T<m_\chi$. The WIMP density falls
exponentially as a result of their annihilation into lighter particles. When
the expansion of the Universe has reduced their density to the point where
annihilation is no longer possible, a relic density ``freezes out'' which
determines the abundance of WIMPs today. This density is just determined by the
annihilation cross section; for a larger cross section freeze-out is delayed
resulting in a lower abundance today and vice versa. The scenario is sketched
in Fig.~1 where the density of WIMPs (in the comoving frame) is shown as a
function of time parametrized as the inverse of the temperature $m_\chi/T$.

\begin{figure}[h]
\centering

\epsfxsize=4in\hspace{0in}\epsffile{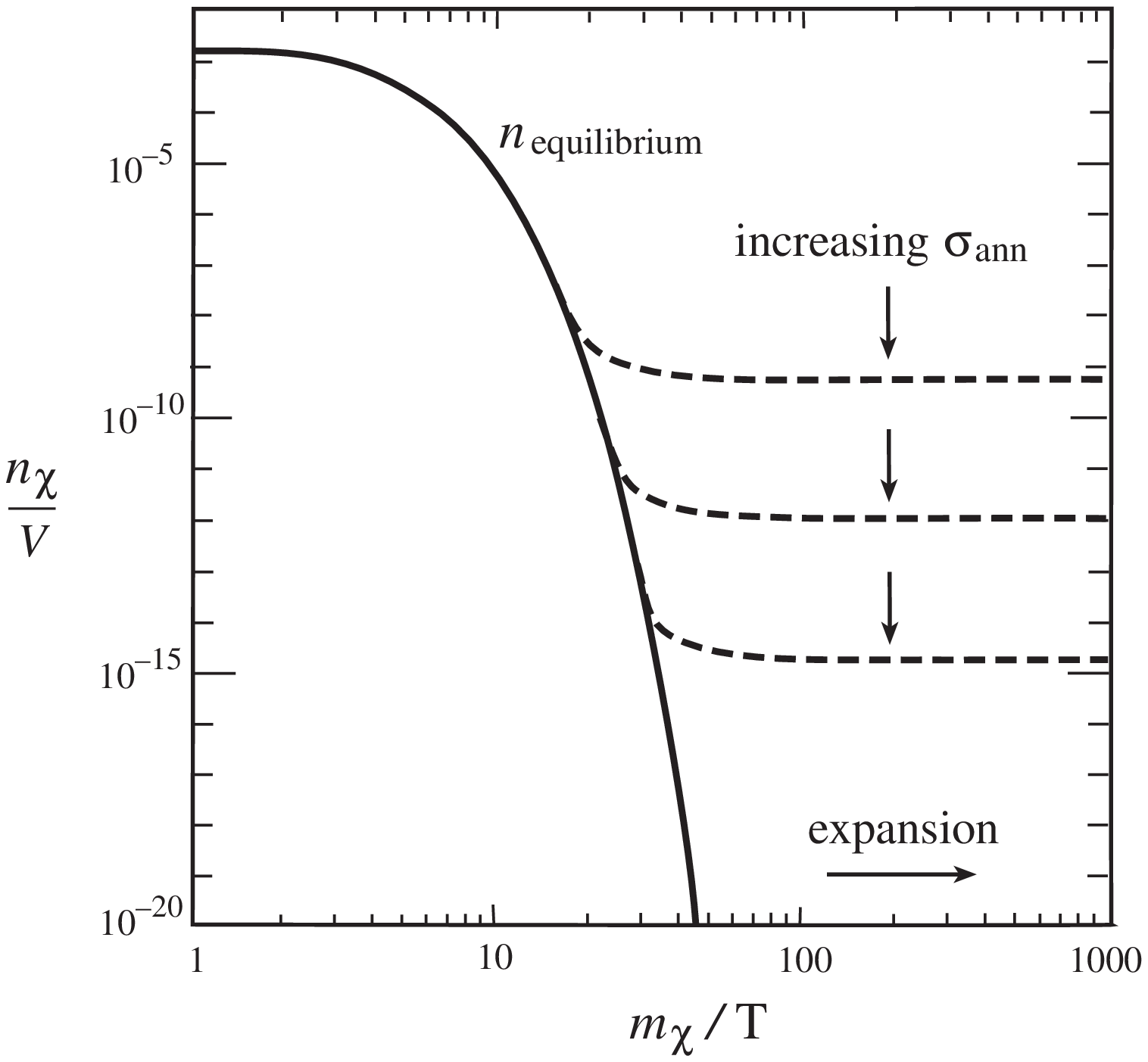}

\medskip
\footnotesize \hspace*{.5in}Fig.~1
\end{figure}

For WIMPs to make up a large fraction of the Universe today, i.e.\ a large
fraction of $\Omega$, their annihilation cross section has to be ``just
right''. The annihilation cross section can be dimensionally written as $
\alpha^2/m_\chi^2$, where $\alpha$ is the fine-structure constant. It then
follows that
\begin{equation}
	\Omega\propto1/\sigma\propto m_\chi^2 \,. \label{omega}
\end{equation}
The critical point is that for $\Omega\simeq1$ we find that $m_\chi\simeq m_W$,
the mass of the weak intermediate boson. There is a deep connection between
critical cosmological density and the weak scale. Weakly interacting particles
which constitute the bulk of the mass of the Universe remain to be discovered.
It may not be an accident that the unruly behavior of radiative corrections in
the Standard Model also requires the existence of such (supersymmetric?)
particles.

When our galaxy was formed the cold dark matter inevitably clustered with the
luminous matter to form a sizeable fraction of the
\begin{equation}
        \rho_{\chi}=0.4\rm~GeV/cm^3  \label{density}
\end{equation}
galactic matter density implied by observed rotation curves. Unlike the
baryons, the dissipationless WIMPs fill the galactic halo which is believed to
be an isothermal sphere of WIMPs with average velocity
\begin{equation}
         v_{\chi}=300\rm\ km/sec \,. \label{velocity}
\end{equation}

In summary, we know everything about these particles (except whether they
really exist!). We know that their mass is of order of the weak boson mass; we
know that they interact weakly. We also know their density and average velocity
in our Galaxy given the assumption that they constitute the dominant component
of the density of our galactic halo as measured by rotation curves.

For a first look at the experimental problem of how to detect these particles
it is sufficient to recall that they are weakly interacting with masses in the
range
\begin{equation}
       \mbox{tens of GeV} < m_{\chi} < \rm several\ TeV \,. \label{GT}
\end{equation}
WIMPs have a mass of order the weak boson mass, in the tens of GeV to several
TeV range. Lower masses are excluded by accelerator and (in)direct searches
with existing detectors while masses beyond several TeV are excluded by
cosmological considerations. Two general techniques, referred to as direct (D)
and indirect (ID),  are pursued to demonstrate the existence of
WIMPs\cite{Seckel}. In direct detectors one observes the energy deposited when
WIMPs elastically scatter off nuclei. The indirect method infers the existence
of WIMPs from observation of their annihilation products. WIMPs will annihilate
into neutrinos; massive WIMPs will annihilate into high-energy neutrinos which
can be detected in high-energy neutrino telescopes\cite{Gaisser}. Throughout
this paper we will assume that such neutrinos are detected in a generic
Cherenkov detector which measures the direction and, to some extent, the energy
of a secondary muon produced by a neutrino of WIMP origin in or near the
instrument. It can oalso detect the showers initiated by electron-neutrinos.

The indirect detection is greatly facilitated by the fact that the sun
represents a dense and nearby source of accumulated cold dark matter
particles\cite{Drees}. Galactic WIMPs, scattering off nuclei in the sun, lose
energy. They may fall below escape velocity and be gravitationally trapped.
Trapped WIMPs eventually come to equilibrium temperature and accumulate near
the center of the sun. While the WIMP density builds up, their annihilation
rate into lighter particles increases until equilibrium is achieved where the
annihilation rate equals half of the capture rate. The sun has thus become a
reservoir of WIMPs which we expect to annihilate mostly into heavy quarks and,
for the heavier WIMPs, into weak bosons. The leptonic decays of the heavy quark
and weak boson annihilation products turn the sun into a source of high-energy
neutrinos with energies in the GeV to TeV range, rather than in the keV to MeV
range typical for neutrinos from thermonuclear burning.

The performance of future detectors is determined by the rate of elastic
scattering of WIMPs in a low-background, germanium detector and, for the
indirect method, by the flux of solar neutrinos of WIMP origin. Both are a
function of WIMP mass and of their elastic cross section on nucleons. In
standard cosmology WIMP capture and annihilation interactions are weak, and we
will suggest that, given this constraint, dimensional analysis is sufficient to
compute the scattering rates in germanium detectors as well as the neutrino
flux from the measured WIMP density in our galactic halo. We derive and compare
rates for direct and indirect detection of weakly interacting particles with
mass $m_\chi \simeq m_W$ assuming

\begin{enumerate}

\item
that WIMPs represent the major fraction of the measured halo density, i.e.
\begin{equation}
\phi_\chi = n_\chi v_\chi = {0.4\over m_\chi} \, {\rm {GeV\over cm^3} \
3\times10^6 {cm\over s} } = {1.2\times10^7\over m_{\chi\rm\,GeV}} \,\rm
cm^{-2} s^{-1} \;,
\label{phi chi}
\end{equation}
where $m_{\chi\rm\,GeV} \equiv (m_\chi/$1~GeV) is in GeV units.

\item
a WIMP-nucleon interaction cross section based on dimensional analysis
\begin{equation}
\sigma(\chi N) = \left(G_F m_N^2\right)^2 {1\over m_W^2} \equiv \sigma_{\rm DA}
= 6\times10^{-42}\rm\,cm^2 \;.
\label{sigma chi N}
\end{equation}

\item
that WIMPs annihilate 10\% of the time in neutrinos (this is just the leptonic
branching ratio of the final state particles in the dominant annihilation
channels $\chi\bar\chi \to W^+W^-$ or $Q\bar Q$, where $Q$ is a heavy quark).

\end{enumerate}

Clearly the cross section for the interaction of WIMPs with matter is
uncertain. Arguments can be invoked to raise or decrease it. Important points
are that i) our choice represents a typical intermediate value, ii) all our
results for event rates scale linearly in the cross section and can be easily
reinterpreted, and iii) the comparison of direct and indirect event rates is
independent of the choice.

Our conclusions will not be surprising\cite{Kamionkowski}. We find that the
direct method is superior if the WIMP interacts coherently and, if its mass is
lower or comparable to the weak boson mass $m_W$. In all other cases, i.e.\ for
relatively heavy WIMPs and for all WIMPs interacting incoherently, the indirect
method is competitive or superior, but it is, of course, held hostage to the
successful deployment of high energy neutrino telescopes with effective area in
the $\sim10^4$--$10^6$~m$^2$ range and with appropriately low threshold.
Especially for heavier WIMPs the indirect technique is powerful because
underground high energy neutrino detectors have been optimized to be sensitive
in the energy region where the neutrino interaction cross section and the range
of the muon are large. A kilometer-size detector probes WIMP masses up to the
TeV-range, beyond which they are excluded by cosmological considerations.

For high energy neutrinos the muon and neutrino are aligned, with good angular
resolution, along a direction pointing back to the sun. The number of
background events of atmospheric neutrino origin in the pixel containing the
signal will be small. The angular spread of secondary muons from neutrinos
coming from the direction of the sun is well described by the
relation\cite{Gaisser} $\sim 1.2^\circ \Big/ \sqrt{E_\mu(\rm TeV)}$.
Measurement of muon energy, which may be only up to order of magnitude accuracy
in some experiments, can be used to infer the WIMP mass from the angular spread
of the signal. The spread contains information on the neutrino energy and,
therefore, the WIMP mass. More realistically, measurement of the muon energy
can be used to reduce the search window around the sun, resulting in a reduced
background.

Our analysis will quantify all statements above in a simple and totally
transparent framework. It finesses all detailed dynamics and gives answers that
are sufficiently accurate considering that the mass of the particle has not
been pinned down.

Before proceeding, we comment on our ansatz for the elastic WIMP-nucleon
scattering cross section. The simplest dimensional analysis implies that the
cross section is $G_F^2 m_N^2$. This correctly describes the $Z$-exchange
diagram of Fig~2a, which is of the form
\begin{equation}
\sigma\ \sim\ G_F^2 {m_N^2m_\chi^2\over (m_N+m_\chi)^2} \;.
\end{equation}

For coherent interactions, which we will emphasize throughout this paper, there
is an additional suppression factor associated with the exchange of the Higgs
particle with a mass of order of the weak boson mass; see Fig~2b. In the
specific diagram shown the Higgs interacts with the heavy quarks in the gluon
condensate associated with the nucleon target. It is of the form
\begin{equation}
\sigma \sim G_F g_H^2 {m_N^2 m_\chi^2\over(m_N+m_\chi)^2} {1\over m_W^2} \;,
\end{equation}
where $g_H \sim \sqrt{G_F}\, m_N$  describes the condensate. Conservatively, we
will use the suppressed WIMP interaction cross section which is appropriate for
coherent scattering.

\begin{figure}[h]
\centering

\epsfxsize=3.25in\hspace{0in}\epsffile{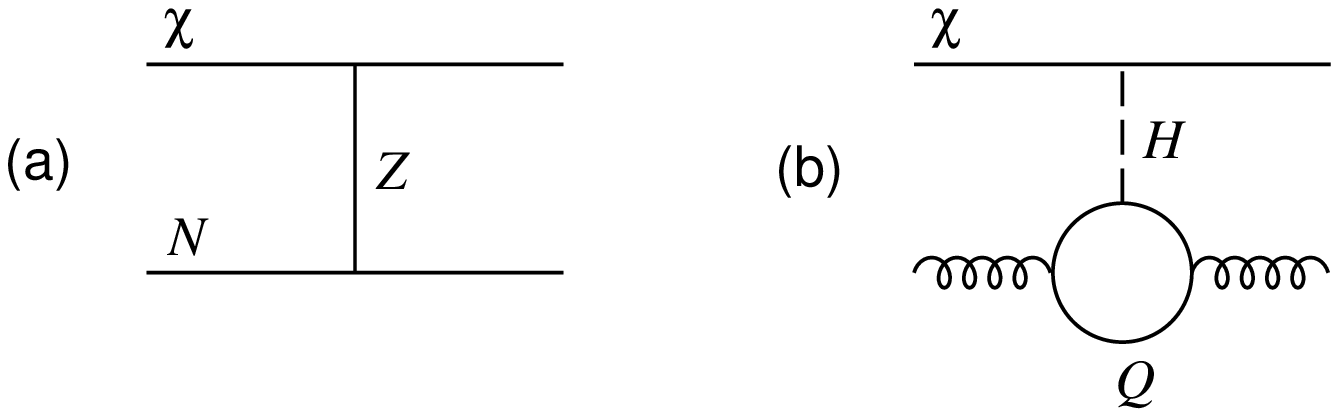}

\medskip
\parbox{6in}{\footnotesize\baselineskip13pt Fig.~2. Examples of (a) incoherent
and (b)~coherent WIMP-nucleon interactions. In (b) the gluon is a constituent
of the target nucleon and $Q$ is a heavy quark.}

\end{figure}

\section{Indirect Detection (ID)}

The number of solar neutrinos of WIMP origin can be calculated in 5 easy steps
by determining

\begin{itemize}

\item
the capture cross section in the sun, which is given by the product of the
number of target nucleons in the sun and the elastic scattering cross section
\begin{equation}
\sigma_\odot = f  \left[ 1.2\times10^{57} \right] \sigma_{\rm DA} \;.
\label{sigma sun}
\end{equation}
This includes a focussing factor $f$ given, as usual, by the ratio of kinetic
and potential energy of the WIMP near the sun. It enhances the capture rate by
a factor 10.

\item
the WIMP flux from the sun which is given by
\begin{equation}
\phi_\odot = \phi_\chi \sigma_\odot / 4\pi d^2 \;,
\label{phi sun}
\end{equation}
where $d=1\rm~a.u. = 1.5\times10^{13}\,cm$.

\item
the actual neutrino flux, which is obtained after inclusion of the branching
ratio. From (\ref{phi chi}),(\ref{sigma chi N}) and (\ref{sigma sun}),(\ref{phi
sun})
\begin{equation}
\phi_\nu = 10^{-1} \times \phi_\odot
= {3\times10^{-5}\over m_{\chi\rm\,GeV}}\rm \, cm^{-2} \, s^{-1} \;.
\end{equation}

\item
the probability to detect the neutrino\cite{Gaisser}, which is proportional to
\begin{eqnarray}
&&P = \rho\sigma_\nu R_\mu,\rm\ with\nonumber\\
&&\rho = \mbox{Avagadro\,\#} = 6\times10^{23}\nonumber\\
&&\sigma_\nu = \mbox{neutrino interaction cross section} = 0.5\times10^{-38}\
E_\nu\rm (GeV)\ cm^{2}\nonumber\\
&&R_\mu = \mbox{muon range} = 500{\rm\ cm}\ E_\mu\rm(GeV)\, \nonumber\\
&&\quad \rm or\nonumber\\
&&P = 2\times 10^{-13} \, m_{\chi\rm\,GeV}^2
\end{eqnarray}
Here we assumed the kinematics of the decay chain
\begin{eqnarray}
\chi\bar\chi &\to& W^+W^- \nonumber \\  \noalign{\vskip-1ex}
             &   & \hspace{2em} \raise1ex\hbox{$\vert$}\!{\rightarrow}
\mu\nu_\mu \nonumber
\end{eqnarray}
with $E_\nu = {1\over2}m_\chi$ (this would be ${1\over3}m_\chi$ for $Q$ decay)
and $E_\mu = {1\over2}E_\nu = {m_\chi\over 4}$.

\item
finally, $dN_{\rm ID} / dA = \phi_{\nu} P = 1.8\times10^{-6} \,
m_{\chi\rm\,GeV} \, \rm\ (year)^{-1} \, (m^2)^{-1}$
\stepcounter{equation}\hfill(\theequation)\break
where $dN_{\rm ID} / dA$ represents the number of events from the sun per unit
area (m$^2$) detected by a neutrino telescope.

\end{itemize}

\noindent
The linear rise of $\sigma_\nu,\, R_\mu$ with energy, which are the origin of
the good detection capability of neutrino telescopes for large WIMP masses, are
valid approximations up to
\begin{eqnarray}
E_\nu &\simeq& {m_\chi\over2}\ \gsim \ {m_W^2\over m_N}\,, \rm\ and\nonumber\\
E_\mu &\simeq& {m_\chi\over4}\ \gsim \ 500\rm\ GeV, \nonumber
\end{eqnarray}
so the approximations are valid for $m_\chi$ well into the TeV mass range. This
is sufficient as $m_\chi\gg 1$~TeV is cosmologically unacceptable.

\section{Direct Detection}

The event rate in a direct detector is proportional to the WIMP cross section,
flux and the density of targets $m_N^{-1}$, i.e.
\begin{equation}
{dN_{\rm D} \over dM} = {1\over m_N} \phi_\chi \sigma_{\rm DA},
\end{equation}
where ${dN_{\rm D}\over dM}$ represents the number of direct events per unit of
target mass.

We can now summarize our results so far by comparing a $10^4$~m$^2$ neutrino
detector, an area typical of the instruments now being deployed, with a
kilogram of hydrogen:
\begin{eqnarray}
dN_{\rm ID}/ dA &=& 1.8\times10^{-2} m_{\chi\rm\,GeV} \rm\ (10^4\,m^2)^{-1}
(year)^{-1} \nonumber\\
dN_{\rm D}/ dM &=& {1.4\over m_{\chi\rm\,GeV}} \rm\ (kg)^{-1} \, (year)^{-1}
\nonumber\\
{dN_{\rm D}/dM\over dN_{\rm ID}/dA} \left(10^4\rm\,m^2\over\rm kg\right)
&=& {7.8\times10^1\over m_{\chi\rm\,GeV}^2}  \label{D/ID}
\end{eqnarray}
Direct detection is superior only in the mass range $m_\chi<10$~GeV, but this
region is, arguably, ruled out by previous searches. Indirect detection is the
preferred technique. This straightforward conclusion may, however,  be
invalidated when WIMPs interact coherently and targets other than hydrogen are
considered. We discuss this next.

\section{Coherent Nuclear Enhancements}

The nuclear dependence of the event rates resides in

\begin{itemize}

\item
the target density factor $m_N^{-1}$ in Eq.~(14). The mass of the target
nucleus $m_A = Am_N$ is substituted for $m_N$.

\item
the coherent enhancement factor ``$A^2$",

\item
the nuclear dependence of the cross section is obtained by the substitution
\begin{eqnarray}
\noalign{\qquad\underline{\rm incoherent}}
\sigma &\sim& G_F^2 {m_N^2m_\chi^2\over (m_N+m_\chi)^2} \to G_F^2  {(Am_N)^2
m_\chi^2\over (Am_N+m_\chi)^2} \nonumber\\ 
\noalign{\qquad\underline{\rm coherent}}
\sigma &\sim& G_F^2 g_H^2 {m_N^2 m_\chi^2\over(m_N+m_\chi)^2} {1\over m_W^2}
\to G_F\left(g_H\over m_W\right)^2 {(Am_N)^2 m_\chi^2\over (Am_N+m_\chi)^2} A^2
\nonumber
\end{eqnarray}
The coherent enhancement factor for a nucleus $A$, including a factor $A^{-1}$
for the target density,  is given by
\begin{equation}
{1\over A} {A^2 (Am_N)^2 m_\chi^2 \over (Am_N+m_\chi)^2 } \,
{(m_N+m_\chi)^2\over m_N^2 m_\chi^2}
= A^3 {(m_N+m_\chi)^2\over (Am_N+m_\chi)^2}
= A^3\left[ 1+{m_\chi\over m_N} \over A+{m_\chi\over m_N} \right]^2 .
\end{equation}

\end{itemize}

Below we list a sample of enhancement factors having in mind germanium instead
of hydrogen for direct detection and oxygen or iron nuclei for capture in the
sun. Their solar abundance relative to hydrogen are 1.1 and 0.2 percent,
respectively.

\begin{table}[h]
\tcaption{Nuclear Enhancement Factors}
\vspace{-.2in}
\[ \arraycolsep=1.5em
\begin{array}{c|ccc}
\hline
& m_\chi = 50 & m_\chi = 500 & m_\chi = 2000\\
\hline
{\rm Ge}\ (A=76)& 7.7\times10^4 & 3.9\times 10^5 & 4.2\times10^5\\
{\rm O}\ (A=16)& 2.5\times10^3 & 3.9\times10^3 & 4.1\times10^3\\
{\rm Fe}\ (A=56)& 4.3\times10^4 & 1.4\times10^5 & 1.7\times10^5\\
\hline
\end{array}
\]
\end{table}

Because of the complications associated with nuclear form factors, calculations
considering just oxygen or iron capture in the sun bracket detailed
computations\cite{Kamionkowski}.

\section{Event Rates for WIMPs with Coherent Interactions}

After inclusion of above coherence factors in Eq.~(\ref{D/ID}) we obtain the
data rates listed below:

\begin{table}[h]
\tcaption{Event Rates per year for Direct and Indirect Detection}
\vspace{-.2in}
\[ \arraycolsep=1.25em
\begin{array}{c|ccc}
\hline
& \rm ID/10^4\,m^2& \rm D/kg& \mbox{D/ID ratio (kg/10$^4$\,m$^2$)}\\
\hline
m_\chi=50& (\rm H)\ 9\times10^{-1}& (\rm H)\
2.8\times10^{-2}&3.1\times10^{-2}\\
& (\rm O)\ 2.5\times10^1 & (\rm Ge)\ 2.2\times10^3& 8.8\times10^1\\
\hline
m_\chi=500& (\rm H)\ 9 & (\rm H)\ 2.8\times10^{-3} & 3\times10^{-4}\\
& (\rm O)\ 4\times10^2 & (\rm Ge)\ 1.1\times10^3& 2.7\\
\hline
m_\chi=2000 & (\rm H)\ 3.6\times10^1 & (\rm H)\ 7\times10^{-4} &
2\times10^{-5}\\
& (\rm O)\ 1.6\times10^3 & (\rm Ge)\ 2.9\times10^2 & 1.8\times10^{-1}\\
\hline
\end{array}
\]
\end{table}

We here assumed a direct detector made of germanium.  Conservatively, only
oxygen, the dominant element in the solar capture rate, was considered in
calculating the indirect rates. We chose 50~GeV and 2~TeV WIMP as illustrative
masses in order to bracket the appropriate range with an illustrative central
value of 500~GeV. One should keep in mind that a 500~GeV WIMP is well out of
reach of present as well as future accelerator searches.

The ratio of direct to indirect events, which is independent of the
WIMP-nucleon cross section, can be summarised by the following equation:
\begin{equation}
{\rm {D\over ID}} = {dN_{\rm D}/dM\over dN_{\rm ID}/dA} \simeq
{7.8\times10^1\over m_{\chi\rm\,GeV}^2} \, {N(A_{\rm D})\over N(A_{\rm ID})
\left[\rho(A_{\rm ID}) / \rho(H) \right]}
\end{equation}
with
\begin{equation}
N(A) \equiv A^3 \left[ 1 + {m_\chi\over m_N} \over A + {m_\chi\over m_N}
\right]^2 \;.
\end{equation}
As in Eq.~(\ref{D/ID}) the units are $10^4\rm\,m^2\over\rm kg$. $A_{\rm D}$ and
$A_{\rm ID}$ are the atomic numbers appropriate for the nuclei involved in the
direct detection and capture in the sun, respectively. The latter is weighted
by its relative mass abundance $\left[\rho(A_{\rm ID}) / \rho(H) \right]$ in
the sun and a summation over elements is understood. This relative evaluation
of the two experimental techniques is in very good qualitative agreement with a
similar analysis performed in the context of supersymmetry\cite{Kamionkowski}.

\section{Final Event Rates}

Our simple evaluations, made so far, overestimate the indirect rates for very
heavy WIMPS because high energy neutrinos, created by annihilation near the
core, may be absorbed in the sun. Absorption is stronger for neutrinos and,
therfore, mostly antineutrinos form the signature for very heavy WIMPS. The
probability that an antineutrino escapes without absorption is well
parametrized by $(1+ 3.8 \times 10^{-4} E_{\nu})^{-7}$, where $E_{\nu} \simeq
m_{\chi}/2$. The final rates for indirect detection are
\begin{equation}
dN_{\rm ID} /dA \simeq \left\{ 1.8\times10^{-2}m_{\chi\rm\,GeV} \right\}
\left\{ 0.011 A^3 \left[ 1+{m_\chi\over m_N}\over A+{m_\chi\over m_N} \right]^2
\right\} \left\{ 1+1.9\times10^{-4} m_{\chi\rm\,GeV} \right\}^{-7} \;.
\end{equation}

The relative merits of the two methods are summarised in the following table,
which establishes that a kilogram of germanium and a $10^4$~m$^2$ are
competitive.

\begin{table}[h]
\tcaption{Event rates and signal to background $(N/B)$.}
\centering
\tabcolsep=1.5em
\begin{tabular}{c|c@{\quad}cc@{\quad}c}
\hline
$m_\chi$ (GeV)&\multicolumn{2}{c}{Direct (/kg/year)}&
\multicolumn{2}{c}{Indirect (/$10^4$\,m$^2$/year)}\\
\hline
& events& $N/B$& events& $N/B$\\
50 & $2.2\times10^3$& 7& $2.3\times10^1$& $\simeq\,1$\\
500 & $1.1\times10^3$& 7& $2\times10^2$& $\simeq\,10^2$\\
2000 & $2.9\times10^2$& 1& $1.7\times10^2$& $\simeq\,10^4$\\
\hline
\end{tabular}
\end{table}

At the lower energy the event rates for the indirect method are underestimated
because also the Earth is a source of neutrinos of WIMP origin.

We conclude that the direct method yields more events for the lower masses,
even when compared to a $10^6$~m$^2$ detector. As expected, the indirect method
is competitive for heavier WIMPs with a detection rate growing like $E_\nu^2$
or $m_\chi^2$. A $10^5$~m$^2$ covers the full WIMP mass range, even if the
WIMPs do not coherently interact with nuclei in the sun; see Table 2. These
conclusions are reinforced after considering the signal-to-noise for both
techniques. We show this in the next section. A summary of our results is shown
in Fig.~3.

\bigskip

\begin{figure}[h]
\centering
\epsfxsize=4.75in\hspace{0in}\epsffile{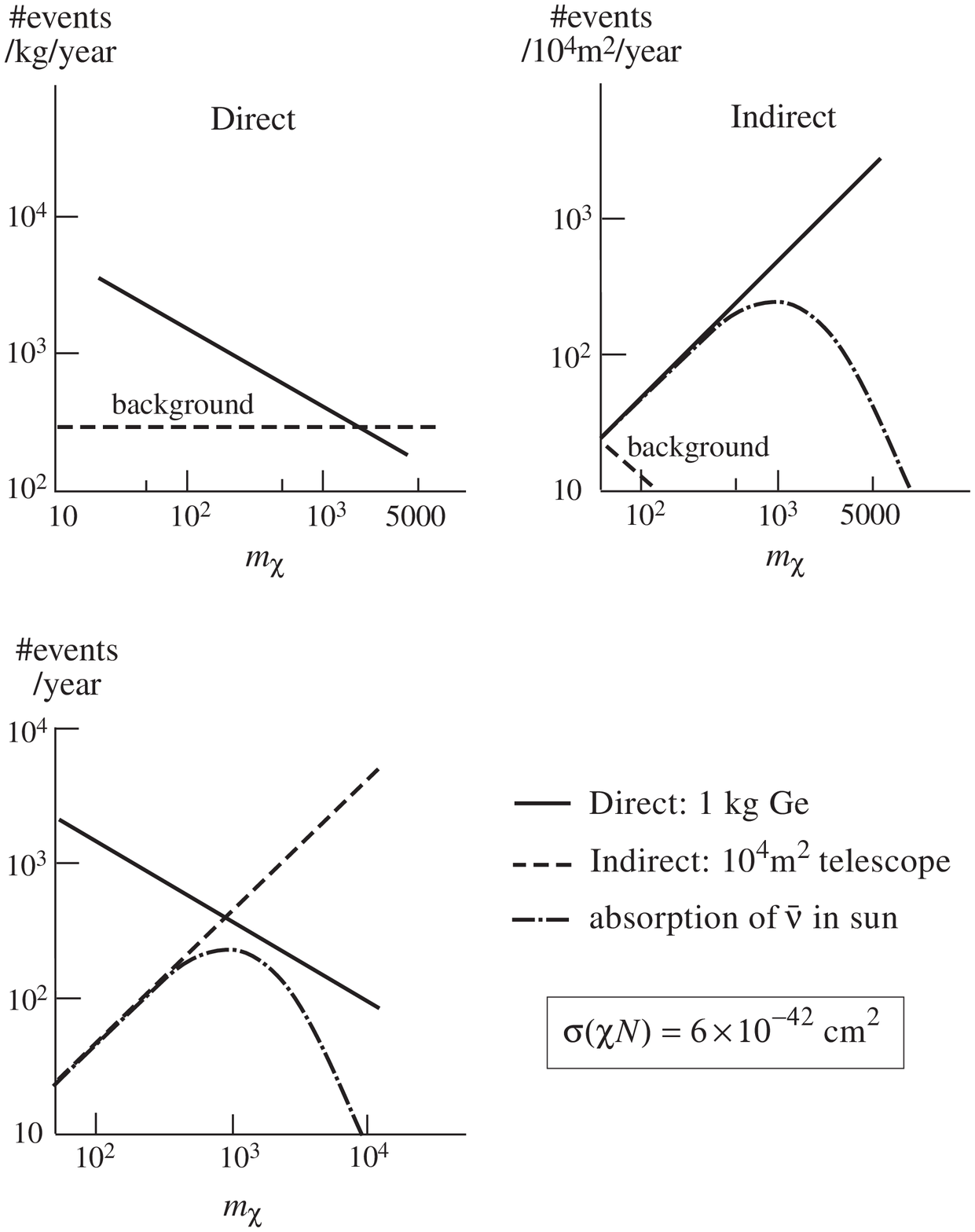}

\bigskip

\parbox{6in}{\footnotesize\baselineskip13pt Fig.~3. Summary of the results. The
results shown are for $\sigma(\chi N) = 6\times10^{-42}\rm\,cm^2$. All event
rates scale linearly in $\sigma(\chi N)$. The relative direct and indirect
rates are independent of $\sigma(\chi N)$.}

\end{figure}

\section{Backgrounds}

\noindent
\underline{Indirect Background}.
For the indirect detection the background event rate is determined by the flux
of atmospheric neutrinos in the detector coming from a pixel around the
sun\cite{Gaisser}. The number of events in a $10^4$~m$^2$ detector is $\sim
10^2/E_\mu$(TeV) and the pixel size is determined by the angle between muon and
neutrino $\sim 1.2^\circ \Big/ \sqrt{E_\mu(\rm TeV)}$. Using the kinematics
$E_\mu \simeq m_\chi/4$ we obtain
\[
B_{\rm ID} = { 10^2/E_\mu({\rm TeV}) \over 2\pi \Big/
\left[ 1.2^\circ {\pi\over 180^\circ} \over \sqrt{E_\mu(\rm TeV)} \right]^2}
= {1.1\times10^5\over m_{\chi\rm\,GeV}^2} \mbox{ per 10$^4$\,m$^2$ per year}
\]
This is only valid for large $m_\chi$, i.e.\ for $E_\mu \cong m_\chi/4
>100$~GeV. Without this approximation we obtain

\begin{table}[h]
\tcaption{}
\centering
\tabcolsep=1.25em
\begin{tabular}{c|ccc}
\hline
            &  \# bkgd. events& \# pixels of solar& bkgd. events\\
\noalign{\vskip-1ex}
            & in 10$^4$\,m$^2$&     size in $2\pi$& per 10$^4$\,m$^2$\\
\noalign{\vskip-1ex}
$E_\mu$(GeV) & in $2\pi$       &                   & per pixel, per year \\
\hline
10& 3200& 140& 23\\
100& 1060& $1.4\times10^3$& 0.8\\
1000& 110& $1.4\times10^4$& $8\times10^{-3}$\\
\hline
\end{tabular}
\end{table}

\noindent
For large $m_\chi$ the signal to background ratio is
\[
\left(N\over B\right)_{\rm ID} \equiv {dN_{\rm ID}/dA\over dB_{\rm ID}/dA}
\simeq
7.2\times10^{-6} m_{\chi\rm\,GeV}^3
\]
Clearly, the extremely optimistic predictions for signal-to-noise are unlikely
to survive the realities of experimental physics. One expects, typically, to
measure muon energy only to order-of-magnitude accuracy in the initial
experiments. The energy of showers initiated by electron neutrinos should be
determined to a factor 2. It is not excluded that future, dedicated experiments
may do better. The conclusion that high energy muons pointing at the sun
represents a superb signature, is unlikely to be invalidated.

\smallskip
\noindent
\underline{Direct Background}: about 300 events per year per
kg\cite{Kamionkowski}. Signal-to-noise therefore exceeds unity up to 2~TeV WIMP
mass.

These considerations were used to estimate the signal-to-noise $N/B$ in
Table~2.

\section{Dynamics?}

We emphasize that above considerations are valid for the specific and much
studied example where the lightest supersymmetric particle is Nature's
WIMP\cite{Drees}. Clearly dynamics, which is now defined, can alter our
conclusions, but only in ``conspiratorial" ways. In the favored scenario the
WIMP is the stable neutralino, i.e.\ the lightest state composed of the
supersymmetric partners of the photon, neutral weak boson and the two Higgs
particles:
\begin{equation}
\chi = z_{11} \tilde W_3 + z_{12} \tilde B + z_{13} \tilde H_1 + z_{14} \tilde
H_2 \,.
\end{equation}
In this specific case of supersymmetry the diagrams in Fig.~2a,b are
proportional to
\begin{eqnarray}
&  g_A^2 ( z_{13}^2 - z_{14}^2 ) \rm\quad and &\\
&  (-z_{11} s_\theta + z_{12} c_\theta) ( z_{13}^2 - z_{14}^2 ) \;, &
\end{eqnarray}
where $g_A$ is the axial coupling of the nucleon and $\theta$ the weak angle.
Even though (22) is not completely general, it is clear that the rates can be
suppressed in a scheme where the neutralino is mostly gaugino-like, i.e.\
$z_{13}=z_{14}=0$. Although such scenarios have been suggested, there is no
consensus. Dynamics can, on the other hand, increase rates as well, sometimes
by well over an order of magnitude, over and above the rates obtained from
dimensional analysis in this paper. Our qualitative conclusions are valid, at
least in some average sense, in supersymmetry. Our results do, in fact, closely
trace the supersymmetry prediction of reference~1 for the choice of Higgs
coupling $\alpha_H=1$, in their notation.

We feel that the development of detectors should be guided by an analysis like
ours rather than by dynamics of theories beyond the standard model for which
there is, at present, no experimental guidance.

\bigskip

\leftline{\bf Acknowledgements}
\medskip

This work was done with M.~Drees who refuses to sign the paper because of its
lack of reverence for supersymmetry. We thank S.~Pakvasa and X.~Tata for a
careful reading of the manuscript.
This research was supported in part by the U.S.~Department of Energy under
Grant No.~DE-FG02-95ER40896 and in part by the University of Wisconsin Research
Committee with funds granted by the Wisconsin Alumni Research Foundation.

\nonumsection{References}\unskip

\end{document}